\documentclass[pdflatex,sn-nature]{sn-jnl}

\usepackage{graphicx}%
\usepackage{multirow}%
\usepackage{amsmath,amssymb,amsfonts}%
\usepackage{amsthm}%
\usepackage{mathrsfs}%
\usepackage[title]{appendix}%
\usepackage{xcolor}%
\usepackage{textcomp}%
\usepackage{manyfoot}%
\usepackage{booktabs}%
\usepackage{algorithm}%
\usepackage{algorithmicx}%
\usepackage{algpseudocode}%
\usepackage{listings}%
\usepackage{hyperref}

\theoremstyle{thmstyleone}%
\theoremstyle{thmstyletwo}%

\theoremstyle{thmstylethree}%

\begin{document}

\title[Hybrid Machine Learning and Physics-based Modelling of Pedestrian Pushing Behaviours at Bottlenecks]{Hybrid Machine Learning and Physics-based Modelling of Pedestrian Pushing Behaviours at Bottlenecks}


\author[1]{\fnm{Qiancheng} \sur{Xu}}\email{xuq@szu.edu.cn}

\author[2]{\fnm{Ezel} \sur{\"Usten}}\email{e.uesten@fz-juelich.de}

\author[2,3]{\fnm{Ahmed} \sur{Alia}}\email{a.alia@fz-juelich.de}

\author*[1]{\fnm{Biao} \sur{He}}\email{hebiao@szu.edu.cn}

\author[1]{\fnm{Renzhong} \sur{Guo}}\email{guorz@szu.edu.cn}

\author[2]{\fnm{Mohcine} \sur{Chraibi}}\email{m.chraibi@fz-juelich.de}

\affil*[1]{\orgdiv{Research Institute for Smart Cities, School of Architecture and Urban Planning}, \orgname{Shenzhen University}, \orgaddress{\city{Shenzhen}, \postcode{518060}, \country{PR China}}}

\affil[2]{\orgdiv{Institute for Advanced Simulation}, \orgname{Forschungszentrum J\"ulich}, \orgaddress{\city{J\"ulich}, \postcode{52428}, \country{Germany}}}

\affil[3]{\orgdiv{Department of Information Technology}, \orgname{An-Najah National University}, \orgaddress{\city{Nablus}, \postcode{P4110257}, \country{Palestine}}}

\abstract{
In high-density crowds, close proximity between pedestrians makes the steady state highly vulnerable to disruption by pushing behaviours, potentially leading to serious accidents.
However, the scarcity of experimental data has hindered systematic studies of its mechanisms and accurate modelling.
Using behavioural data from bottleneck experiments, we investigate pedestrian heterogeneity in pushing tendencies, showing that pedestrians tend to push under high-motivation and in wider corridors. 
We introduce a spatial discretization method to encode neighbour states into feature vectors, serving together with pedestrian pushing tendencies as inputs to a random forest model for predicting pushing behaviours.
Through comparing speed-headway relationships, we reveal that pushing behaviours correspond to an aggressive space-utilization movement strategy. 
Consequently, we propose a hybrid machine learning and physics-based model integrating pushing tendencies heterogeneity, pushing behaviours prediction, and dynamic movement strategies adjustment.
Validations show that the hybrid model effectively reproduces experimental crowd dynamics and fits to incorporate additional behaviours.
}

\maketitle

\section{Introduction}\label{sec1}
Crowd accidents are prone to occur in large-scale gatherings, such as sporting events, theatrical performances, and festival celebrations, often resulting in heavy casualties and massive property losses~\cite{sieben2023inside, helbing2007dynamics, zhen2008analysis, illiyas2013human}.
Since 2010, more than 100 crowd accidents have been reported worldwide, causing the loss of at least 4,500 lives~\cite{feliciani2023trends}.
These accidents are often linked to overcrowding due to poor facility layouts and ineffective management strategies.
When the density of the crowd is low, pedestrians have adequate personal space for self-organization, allowing for orderly movement and sufficient response time for emergencies.
However, personal space diminishes as density increases, making the crowd more vulnerable to accidents.
In such high-density scenarios, unfair behaviours, such as pushing, become more disruptive, as they further encroach on the limited personal space of others and can unintentionally cause others to push~\cite{adrian2020crowds, lugering2023psychological}. 
The cascading effects of pushing behaviours can disrupt pedestrian orderly movement and exacerbate the local density, significantly increasing the risk in emergencies~\cite{garcimartin2016flow}.
Understanding the mechanisms of pushing behaviours is essential for developing accurate pedestrian dynamics models that can simulate crowd movement during emergencies and help optimize crowd management strategies.

Pedestrian behaviours are often studied through controlled experiments, such as anticipating in bidirectional flow scenarios~\cite{murakami2021mutual}, detouring in circle antipode scenarios~\cite{xiao2019investigation}, and following in single-file scenarios~\cite{ma2021spontaneous}.
Bottleneck scenarios are particularly suitable for studying pushing behaviours, as they can create high-motivation and high-density conditions that conveniently allow for frequent observations of pushing.
Previous bottleneck experiments have revealed collective phenomena such as the zipper effect~\cite{seyfried2009new}, the clogging~\cite{zuriguel2014clogging}, and the ``faster-is-slower/faster''~\cite{helbing2005self,haghani2019push}.
However, these bottleneck studies were conducted primarily based on head trajectories and wall pressures, which did not focus on collecting the behavioural data required to analyse the mechanisms of pushing behaviours.
The bottleneck experiments conducted at the University of Wuppertal present a significant advance by providing an annotated data set that combines trajectory data with a four-stage classification system to identify the timing and location of pushing behaviours~\cite{adrian2020crowds,usten2022pushing}.
Using the data set, Alia et al. proposed several deep learning approaches to automatically identify pushing behaviours in videos of crowds~\cite{alia2022hybrid, alia2023cloud, alia2024novel}, L\"ugering et al. identified the presence of pushing propagation in crowds~\cite{lugering2023psychological}.
This annotated data set offers a unique opportunity to systematically study the pushing behaviour and develop advanced pedestrian dynamics models.

Microscopic models, which describe pedestrian movements at the individual level, are particularly suitable for incorporating various pedestrian behaviours. 
These models are typically built on a fundamental movement strategy that follows two primary principles: moving toward a target and avoiding neighbours.
To improve realism, these models have integrated additional behaviours such as following~\cite{xie2022experiment}, anticipating~\cite{xu2021anticipation}, detouring~\cite{xu2024analysis}, and grouping~\cite{wu2023force}.
Pushing behaviours have also been incorporated, for example, through position-jump processes in the cellular automata~\cite{yates2015incorporating} or physical forces between agents in the social force model~\cite{helbing2000simulating}.
However, these models typically employ a single, fixed movement strategy, while real-world pedestrians adapt their behaviours dynamically, shifting between different strategies over time, resulting in a heterogeneous composition~\cite{adrian2020crowds,usten2022pushing}.
Consequently, such models constrained to a single strategy have limitations in fully capturing the complexity and adaptability of real-world pedestrian movement.

This study aims to address these limitations by examining the mechanisms underlying pedestrian pushing behaviours using the annotated experimental data~\cite{adrian2020crowds,usten2022pushing}. 
The analysis begins with an assumption that whether a pedestrian engages in pushing behaviours is determined by its internal pushing tendency and external effects from neighbours.
A novel indicator, termed ``free pushing intensity'', is introduced to reflect the heterogeneity among pedestrians in terms of their internal push tendencies.
A spatial discretization method is developed to encode the states of the neighbouring pedestrians into feature vectors, creating a structured and labelled data set that comprises 206,156 non-pushing samples and 93,669 pushing samples.
The generated data set is further enhanced by integrating pedestrian free pushing intensities, which is used to train and validate a random forest algorithm~\cite{breiman2001random} for predicting pushing behaviours.
Building upon this, a hybrid machine learning and physics-based model for pedestrian dynamics is proposed where the distribution of agents' free pushing intensities is derived from experimental data, the behaviours agents engaged in are predicted using the trained random forest classifier, and agents' movement strategies are dynamically adjusted based on their behaviours. 
The proposed hybrid model demonstrates strong performance in high-motivation scenarios where pushing behaviours are frequent and shows promise for further improvement by incorporating additional behaviours.

\section{Results}\label{sec2}
\subsection{Collection of experimental data}
\begin{figure}[htbp]
    \centering
    \includegraphics[width=\linewidth]{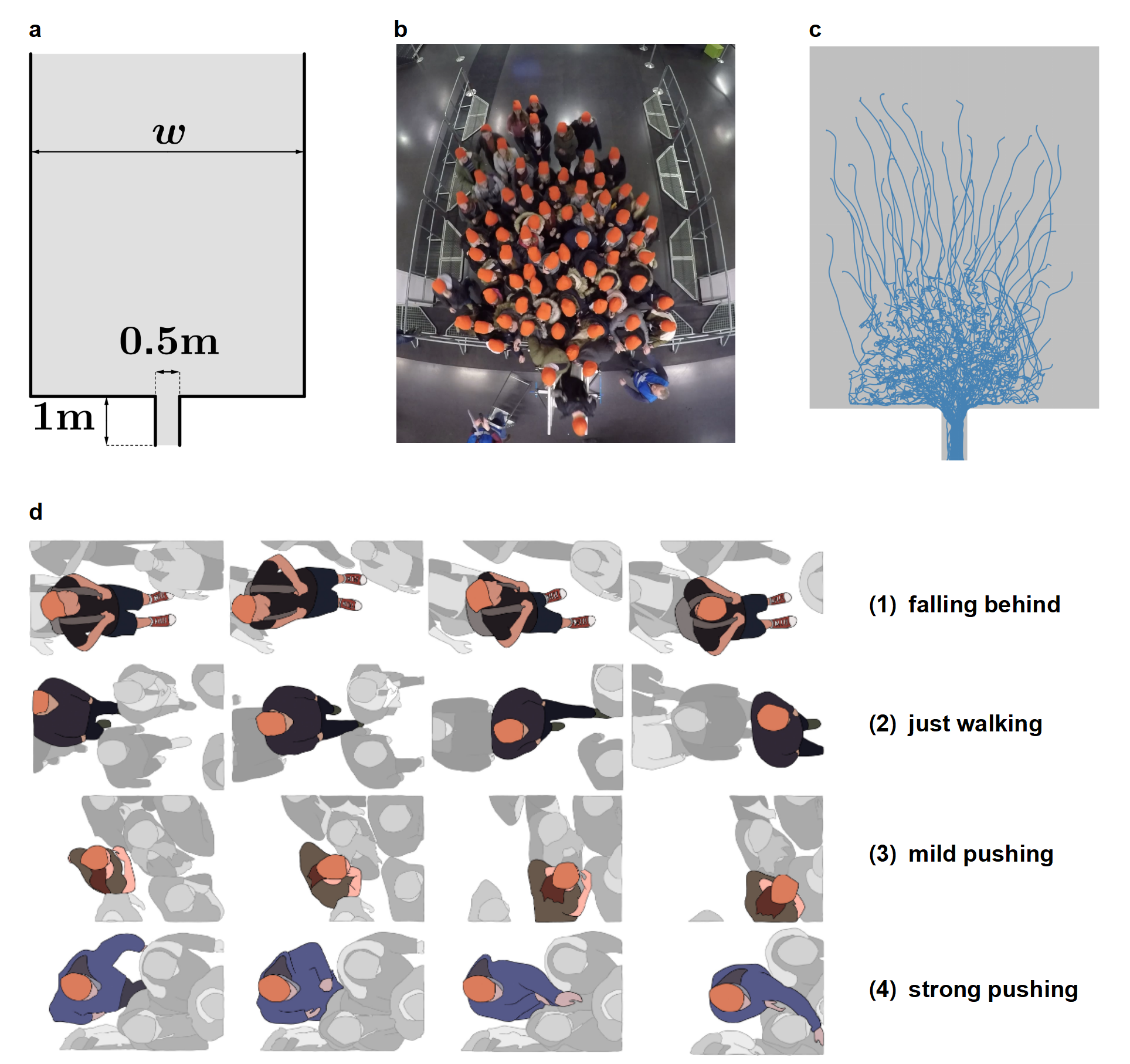}
    \caption{Experimental data collection. 
    a, Experimental structure.
    b, A frame of the recorded experimental video.
    c, Pedestrian head trajectories extracted with PeTrack.
    d, The four-stage category system for pedestrian behaviours, the same illustration from \"Usten et al.~\citep{usten2022pushing} is used.
    } 
    \label{fig:expDatCol}
\end{figure}

The pedestrian experiment was conducted at the University of Wuppertal in January 2018~\citep{adrian2020crowds}. 
The experimental structure is shown in Fig.~\ref{fig:expDatCol}a, a bottleneck scenario with an entrance gate of 0.5 meters.
The width of the corridor leading directly to the entrance, indicated by $w$, was varied and set as 1.2, 3.4, 4.5, and 5.6 meters.
Pedestrians were loosely placed within the corridor and instructed to leave the experimental structure through the entrance gate.
The experiments were carried out twice for each value of $w$, with high- and low-motivation, respectively. 
The motivation level was adjusted by guiding pedestrians with different instructions (see Methods for details).
Experimental processes were recorded with overhead cameras at 25 fps, as shown in Fig.~\ref{fig:expDatCol}b. 
The PeTrack software~\cite{boltes2013collecting} was used to automatically extract the head trajectories of pedestrians from these recorded videos, as shown in Fig.~\ref{fig:expDatCol}c.
Furthermore, with repeated viewing of these recorded videos, trained observers classified pedestrian behaviours at each second according to a four-stage category system~\citep{usten2022pushing}, as shown in Fig.~\ref{fig:expDatCol}d.
The system consists of four categories to annotate all the behaviours that can be seen throughout the experiments, which are (1) falling behind, (2) just walking, (3) mild pushing, and (4) strong pushing. 
Although the ordinal numbers assigned to these four categories were not designed to establish a linear relationship with the pushing intensities, a bigger ordinal number indeed corresponds to a higher pushing intensity of the behaviours in the corresponding category.
Therefore, these ordinal numbers are employed in subsequent analytical procedures to represent the pushing intensities of pedestrians when they engage in the corresponding behaviours.
By integrating the trajectory and behaviour data of pedestrians, 299,825 samples were generated, capturing the pushing intensities of pedestrians at various times and locations.

\subsection{Heterogeneity of pedestrian internal pushing tendencies}
The pushing intensity of a pedestrian is assumed to be determined by its internal pushing tendency and the external effects of neighbouring pedestrians.
The internal pushing tendency of a pedestrian is a concentrated reflection of the pedestrian's physical and psychological attributes.
It is determined at the start of the experiment and remains constant throughout the experimental process.
To quantify the internal pushing tendency of pedestrian $i$, an indicator termed ``free pushing intensity'' (denoted as ${P}_i^0$) is adopted.
It is calculated as the mean value of pedestrian $i$'s pushing intensities from leaving the initial position until reaching the entrance.  
Averaging a pedestrian's pushing intensities throughout an experiment offers a practical representation of the pedestrian's overall pushing tendency during the experiment and helps to capture the heterogeneity of pedestrian internal pushing tendencies.

\begin{figure}[htbp]
    \centering
   \includegraphics[width=\linewidth]{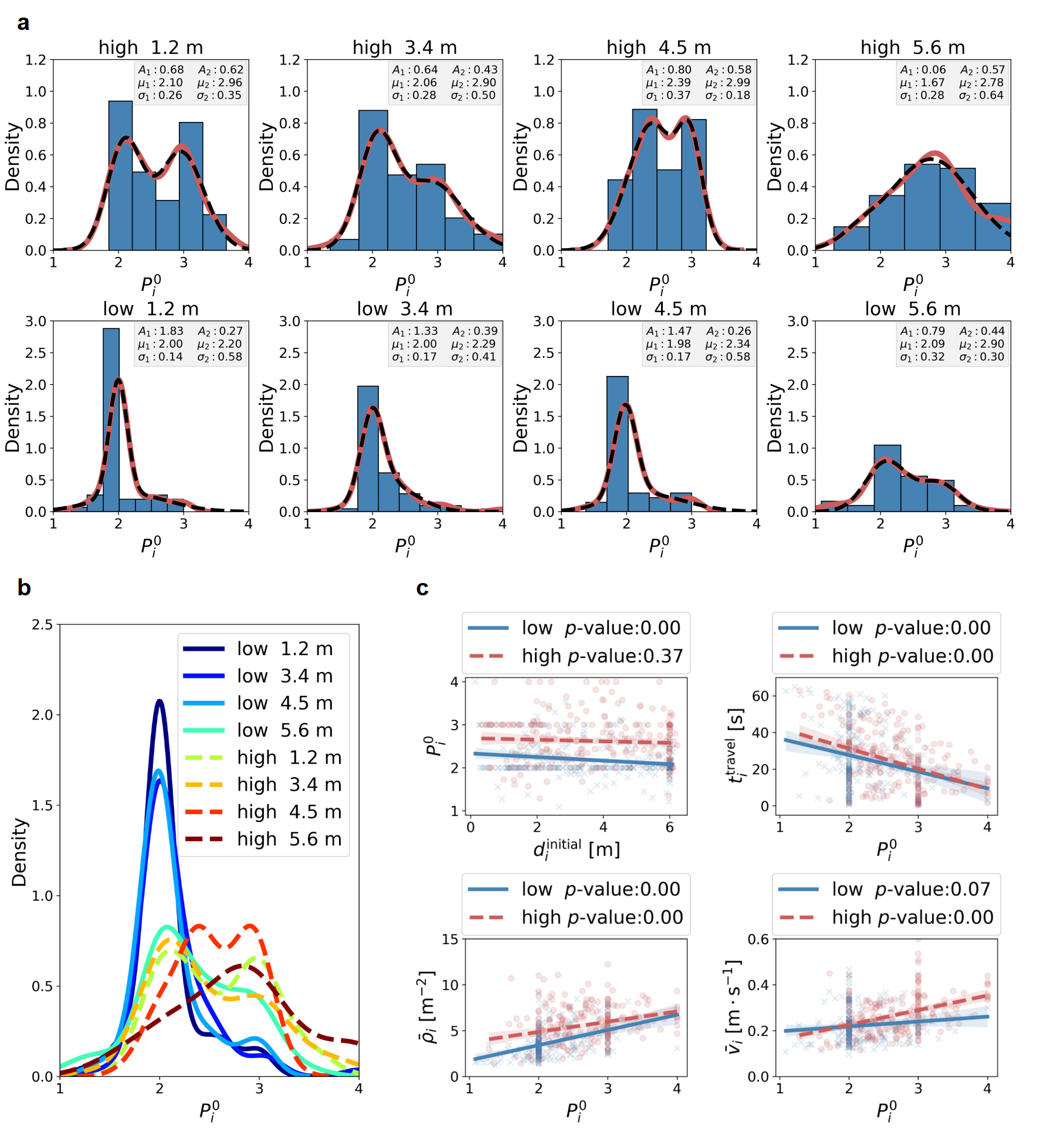}
    \caption{Heterogeneity of pedestrian internal pushing tendencies.
    a, The distributions of pedestrian free pushing intensities, for experiments with different widths of corridor ($w=1.2,~3.4,~4.5,~5.6$ m) and motivation levels (high/low). 
    b, The kernel density estimations of experiments with different widths of corridor and motivation levels.
    c, The relationships between the pairs of variables ($d_i^{\rm{initial}}$, $P_i^0$), ($P_i^0$, $t_i^{\rm{travel}}$), ($P_i^0$, $\bar{\rho}_i$) and ($P_i^0$, $\bar{v}_i$). 
    }
    \label{fig:intFeaAna}
\end{figure}

In order to show the influence of the corridor width and the motivation level on pedestrian internal pushing tendencies in experiments, Fig.~\ref{fig:intFeaAna}a illustrates the distributions of pedestrian free pushing intensities, for experiments with different widths of corridor ($w=1.2,~3.4,~4.5,~5.6$ m) and motivation levels (high/low).
These distributions are depicted using histograms (blue bins) and kernel density estimations (red curves), where Scott's rule determines the number of bins in histograms and the smoothing bandwidths of kernel density estimations.
Since these kernel density estimations show clear bimodal or unimodal distributions, they are fitted by the following function
\begin{equation}
\label{equ:bimodalDist}
f(P_i^0)=A_1 \cdot \exp\left(\frac{-(P_i^0-\mu_1)^2}{2 \cdot \sigma_1^{~2}}\right)+A_2 \cdot \exp\left(\frac{-(P_i^0-\mu_2)^2}{2 \cdot \sigma_2^{~2}}\right),
\end{equation}
where $A_1$, $\mu_1$, $\sigma_1$, $A_2$, $\mu_2$, and $\sigma_2$ are fitting parameters.
The fitting results for experiments with different widths of corridor and motivation levels are shown with black dashed curves in the corresponding subfigures, and the fitting parameters are provided. 
To make a clear comparison, Fig.~\ref{fig:intFeaAna}b collects kernel density estimations of experiments with different widths of corridor and motivation levels.
In low-motivation experiments, an increase in the width of the corridor induces pedestrian free pushing intensities from a unimodal distribution centred at approximately 2 transfer to a bimodal distribution with prominent peaks around 2 and 3.
In contrast, in high-motivation experiments, increasing the width of the corridor facilitates a transition from a bimodal distribution, with peaks at 2 and 3, to a unimodal distribution centred around 3. 
It indicates that pedestrians tend to exhibit a stronger internal pushing tendency in the experiment with a high-motivation level and a wider corridor.

The influence of pedestrian initial positions on their free pushing intensities, as well as the influence of pedestrian free pushing intensities on their dynamics, is examined to validate the suitability of using the free pushing intensity as a measure of pedestrian internal pushing tendencies.
The distance from the initial position to the midpoint of the entrance is calculated for each pedestrian, which is denoted as $d_i^{\rm{initial}}$.
Three additional quantities related to pedestrian dynamics are also calculated, which are the time taken to travel from the initial position to the entrance (denoted as $t_i^{\rm{travel}}$), the mean density (denoted as $\bar{\rho}_i$), and the mean speed (denoted as $\bar{v}_i$) over the period from leaving the initial position until reaching the entrance.
The densities of pedestrians are measured using the Voronoi method~\cite{steffen2010methods}.
According to the motivation levels, experimental data are categorized into two groups (high-/low-motivation).
Linear regressions are then performed within each group to evaluate the relationships between the pairs of variables ($d_i^{\rm{initial}}$, $P_i^0$), ($P_i^0$, $t_i^{\rm{travel}}$), ($P_i^0$, $\bar{\rho}_i$) and ($P_i^0$, $\bar{v}_i$), where the first variable in each pair is the independent variable and the second is the dependent variable. 
The results of these linear regressions and the corresponding $p$ values of the T-tests are shown in Fig.~\ref{fig:intFeaAna}c.
Here, $p$-values smaller than 0.05 indicate that the independent variable exerts a statistically significant effect on the dependent variable.

In high-motivation experiments, the initial distance $d_i^{\rm{initial}}$ does not significantly affect the free pushing intensity $P_i^0$.
Besides, an increase in the free pushing intensity $P_i^0$ is associated with a shorter travel time $t_i^{\rm{travel}}$, a higher mean density $\bar{\rho}_i$, and a higher mean speed $\bar{v}_i$.
However, differences appear in low-motivation experiments, where the free pushing intensity  $P_i^0$ decreases as the initial distance $d_i^{\rm{initial}}$ increases, and $P_i^0$ has no significant effect on the mean speed $\bar{v}_i$.
These differences can be attributed to the experimental instructions.
In high-motivation experiments, pedestrians were informed that entering earlier would bring better benefits, prompting pedestrians at all initial positions to aim for faster entry, which minimized the effect of initial positions on the free pushing intensity.
In contrast, in low-motivation experiments, where benefits were independent of entry order, pedestrians further from the entrance faced a positional disadvantage in reaching it, displayed less urgency and were more likely to have a lower free pushing intensity.
Additionally, the mean speed $\bar{v}_i$ is approximately the initial distance $d_i^{\rm{initial}}$ divided by the travel time $t_i^{\rm{travel}}$.
In high-motivation experiments, the free pushing intensity is not influenced by the initial distance, allowing the free pushing intensity to affect the order of entry greatly. 
As a result, some pedestrians further from the entrance but with a high free pushing intensity can enter more quickly than those with a low free pushing intensity, thus reducing the travel time and increasing the mean speed. 
However, in low-motivation experiments, the free pushing intensity decreases with increasing initial distance, making the initial distance the primary factor in determining entry order rather than the free pushing intensity.
This dependence on the initial distance reduces the influence of the free pushing intensity on the mean speed.
The results of liner regressions confirm that, at least in high-motivation experiments where the pushing behaviours are frequent, the free pushing intensity is a suitable indicator of pedestrian internal pushing tendencies, since it is not affected by pedestrian initial positions and has strong correlations with pedestrian dynamics. 


\subsection{Random forest-based prediction of pushing behaviours}
\begin{figure}[htbp]
    \centering
    \includegraphics[width=\linewidth]{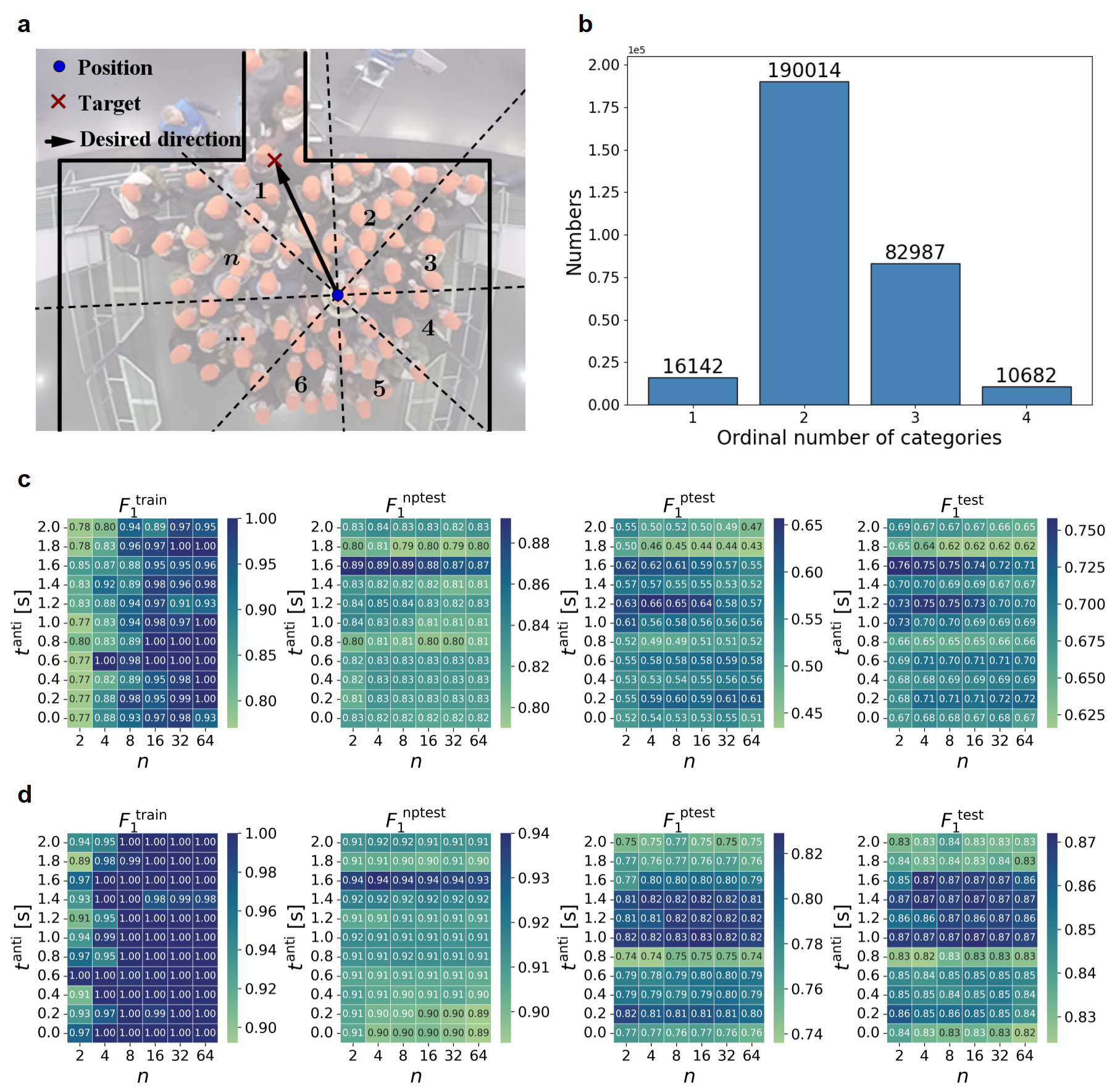}
    \caption{
    Random forest-based prediction of pushing behaviours.
    a, The spatial discretization method for describing the state of neighbours quantitatively.
    b, The number of samples in each category of behaviours.
    c, The performance of the classifiers trained with the first strategy.
    d, The performance of the classifiers trained with the second strategy.
    }
    \label{fig:neigEffe}
\end{figure}

While the free pushing intensity of a pedestrian remains constant during the experiment, its actual pushing intensity fluctuates due to the external effects of neighbouring pedestrians.
To predict pedestrian actual pushing intensities, an efficient machine learning-based classification algorithm is employed. 
The random forest algorithm~\cite{breiman2001random} is selected for its many advantages, including robustness to overfitting, the ability to model nonlinear relationships, and the effectiveness of handling unbalanced data sets. 
The annotated experimental data set undergoes a two-step pre-processing procedure to prepare the required data set for training and evaluation.

a. Describing the state of each pedestrian's neighbours quantitatively with a spatial discretization method illustrated in Fig.~\ref{fig:neigEffe}a.
First, the desired direction of a pedestrian is calculated as a unit vector pointing from its current position to the target, the midpoint of the entrance. 
Next, the space around the pedestrian is divided into $n$ equal regions, each radiating from the position of the pedestrian, with the first region aligned with the desired direction.
The neighbours in each region are described by four features: the closest distance between the pedestrian and the neighbours in region $n$ (denoted as $N^d_n$), the mean effective speed of the neighbours in region $n$ (denoted as $N^v_n$), the mean density of the neighbours in region $n$ (denoted as $N^\rho_n$), and the mean pushing intensity of the neighbours in region $n$ (denoted as $N^p_n$).
The effective speed is defined as the component of a neighbour's velocity in the desired direction of the pedestrian.
Finally, the state of the pedestrian neighbours is represented by a feature vector $[N^d_1, N^v_1, N^\rho_1, N^p_1, \cdots, N^d_n, N^v_n, N^\rho_n, N^p_n]$.

b. Converting the classification task from multi-class to binary.  
Since the number of samples in categories (1) falling behind and (4) strong pushing is significantly smaller compared to in categories (2) just walking and (3) strong pushing (as shown in Fig.~\ref{fig:neigEffe}b), the samples are recategorized into two classes: categories (1) and (2) are grouped as non-pushing, while categories (3) and (4) are grouped as pushing. 
This transformation was also employed by Alia et al.’ studies~\cite{alia2022hybrid, alia2023cloud, alia2024novel}, which has proven generally effective in alleviating the imbalance issue and allowing the creation of a more efficient classifier for pushing behaviours.
Following this transformation, the new data set consists of 206,156 non-pushing samples and 93,669 pushing samples, resulting in a roughly 2:1 ratio, suitable for training the random forest classifier.

The generated data set is then used to train random forest classifiers to predict whether pedestrian behaviours will be pushing or non-pushing.
To evaluate the impact of the pedestrian free pushing intensity $P_i^0$ on classification performance, two training strategies are implemented. 
The first uses only the feature vector representing the state of neighbours as the input, while the second strategy combines this feature vector and $P_i^0$ as the input.
For each strategy, classifiers with different values of $n$ (the number of discrete regions) and anticipation time $t^{\rm{anti}}$ are trained and evaluated, where the anticipation time means using the input at time $t$ to predict the behaviour at time $t + t^{\rm{anti}}$.
To ensure an efficient evaluation, the data set is split by pedestrian IDs into training and test sets with an 80:20 ratio. 
This technique prevents data leakage and ensures that pushing to non-pushing ratio of each set is consistent with the overall data set.
Furthermore, cross-validation is performed for each classifier to tune the random forest parameters, including the number of trees and the maximum depth of trees, aiming to maximize the macro $F_1$ score of the classifier.  
The macro $F_1$ score metric is used to evaluate the performance, as it is particularly valuable for imbalanced classification problems.
It treats the classes of pushing and non-pushing equally by averaging their individual $F_1$ scores~\cite{devries2021using}.
 
Fig.~\ref{fig:neigEffe}c and Fig.~\ref{fig:neigEffe}d display the performance of these classifiers trained with the first and second strategies, respectively.
Subfigures from left to right are the macro $F_1$ score for the training set (denoted as $F_1^{\rm{train}}$), the $F_1$ score for non-pushing samples in the test set (denoted as $F_1^{\rm{nptest}}$), the $F_1$ score for pushing samples in the test set (denoted as $F_1^{\rm{ptest}}$) and the macro $F_1$ score for the test set (denoted as $F_1^{\rm{test}}$).
In the first strategy, the optimal value of $F_1^{\rm{test}}$ is 0.758, which is achieved when $n=2$ and $t^{\rm{anti}}=1.6~\rm{s}$.
For this classifier, the values of $F_1^{\rm{train}}$, $F_1^{\rm{nptest}}$, and $F_1^{\rm{ptest}}$ are 0.855, 0.892, and 0.624 respectively, reflecting high accuracy for non-pushing samples but a lower accuracy for pushing samples.
In the second strategy, which includes pedestrian free pushing intensities as a feature, the optimal value of $F_1^{\rm{test}}$ improves to 0.872 with $n=16$ and $t^{\rm{anti}}=1~\rm{s}$.
Here, the values
of $F_1^{\rm{train}}$, $F_1^{\rm{nptest}}$, and $F_1^{\rm{ptest}}$ increase to 1, 0.914, and 0.829, indicating enhanced predictive performance for both non-pushing and pushing samples.
Furthermore, the prediction performance is better when the anticipation time $t^{\rm{anti}}$ is between 1 and 1.6 seconds. 
This suggests that pedestrians may take approximately 1 to 1.6 seconds to respond with pushing behaviour in reaction to the effect of neighbours.
The performance improvement achieved with the second training strategy confirms the combined influence of pedestrian internal pushing tendencies and the external effects from neighbours on pedestrian behaviours.
In addition, integrating the trained random forest classifier into a physics-based model for pedestrian dynamics is expected to realize accurate predictions of pushing behaviours in simulations.

\subsection{Movement strategies of pushing and non-pushing behaviours}
\begin{figure}[htbp]
    \centering
    \includegraphics[width=\linewidth]{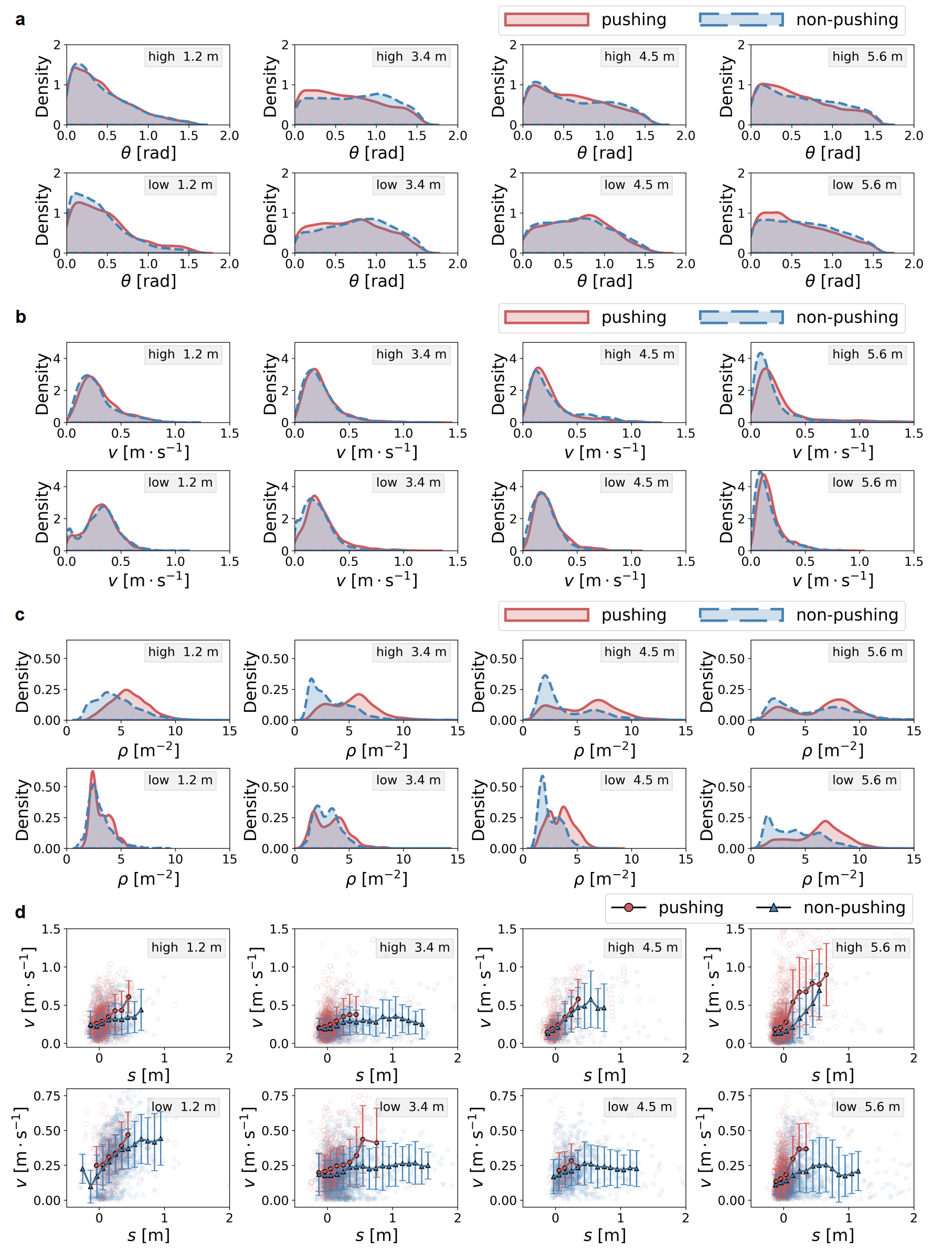}
    \newpage
    \caption{
    Movement strategies of pushing and non-pushing behaviours. 
    a, The probability density function (PDF) of pedestrian deviation angles ($\theta$) during pushing and non-pushing behaviours.
    b, The PDFs of pedestrian speeds ($v$) during pushing and non-pushing behaviours. 
    c, The PDFs of pedestrian densities ($\rho$) during pushing and non-pushing behaviours. 
    d, The relationships between pedestrian available distances in the direction of movement ($s$) and their corresponding speeds ($v$) during pushing and non-pushing behaviours.}
    \label{fig:MoviDiff}
\end{figure}

Pedestrian behaviours are recategorized into pushing and non-pushing groups when training the random forest classifiers.
Here, the differences in movement strategies associated with pushing and non-pushing behaviours are studied.
Pedestrian movement strategies are primarily reflected by adjustments of velocities in response to changes in the state of neighbours, which can be decomposed into adjustments in the direction of movement and the magnitude of speed.
Fig.~\ref{fig:MoviDiff}a presents the probability density function (PDF) of deviation angles (denoted as $\theta$) during the pushing and non-pushing behaviours.
The deviation angle is defined as the absolute angle between a pedestrian's desired direction and its actual direction of movement.
These PDFs are estimated using kernel density estimations, with the smoothing bandwidth determined by Scott's rule.
Although the differences in these PDFs are subtle, non-pushing behaviours exhibit slightly larger deviation angles compared to pushing behaviours when the width of the corridor exceeds 1.2 meters. 
This indicates that in scenarios with sufficient space for directional adjustments, pedestrians engaging in pushing behaviours tend to align more closely with their desired direction than those who engage in non-pushing behaviours.
Next, Fig.~\ref{fig:MoviDiff}b compares the PDFs of pedestrian speeds (denoted as $v$) during pushing and non-pushing behaviours. 
The speed $v$ is calculated as the average speed of a pedestrian over a 10-frame interval, corresponding to 0.4 seconds. 
The distributions of speeds for both behaviours appear to be nearly identical.

However, this observed similarity between pushing and non-pushing behaviours in distributions of deviation angles and speeds does not necessarily imply identical movement strategies.
Fig.~\ref{fig:MoviDiff}c provides additional information by comparing the PDFs of pedestrian densities (denoted as $\rho$) during pushing and non-pushing behaviours. 
The results reveal a marked difference: pedestrians engaging in pushing behaviours are more likely to move in higher-density situations than non-pushing behaviours.
This highlights a key divergence in movement strategies: pedestrians who engage in pushing maintain a similar magnitude of speed compared to non-pushing pedestrians, even in situations that are more crowded than those experienced by non-pushing pedestrians.
To further validate this divergence, Fig.~\ref{fig:MoviDiff}d examines the relationships between pedestrian available distances in the direction of movement (denoted as $s$) and their speeds $v$ during pushing and non-pushing behaviours, also known as the speed-headway relationship. 
The available distance $s$ is defined as the maximum distance that a pedestrian can move in its direction of movement without overlapping with the area occupied by other pedestrians, where the area of each pedestrian is approximated as a circular disk with a radius of 0.18 meters.  
To illustrate how the speed $v$ changes with the available distance $s$, the experimental data is divided into bins of 0.1 meters, and the average speed for each bin is calculated. 
The relationships between the available distance and the speed reveal two significant distinctions.
First, pedestrians who engage in pushing behaviours generally have a shorter available distance compared to those engaging in non-pushing behaviours, indicating closer proximity to their front neighbours.
Second, for the same available distance, pedestrians who engage in pushing achieve higher speeds than those who do not. 
These observations confirm that pushing behaviours correspond to a more aggressive space-utilization movement strategy compared to non-pushing behaviours.

\subsection{Hybrid machine learning and physics-based model}
\begin{figure}[htbp]
    \centering
    \includegraphics[width=\linewidth]{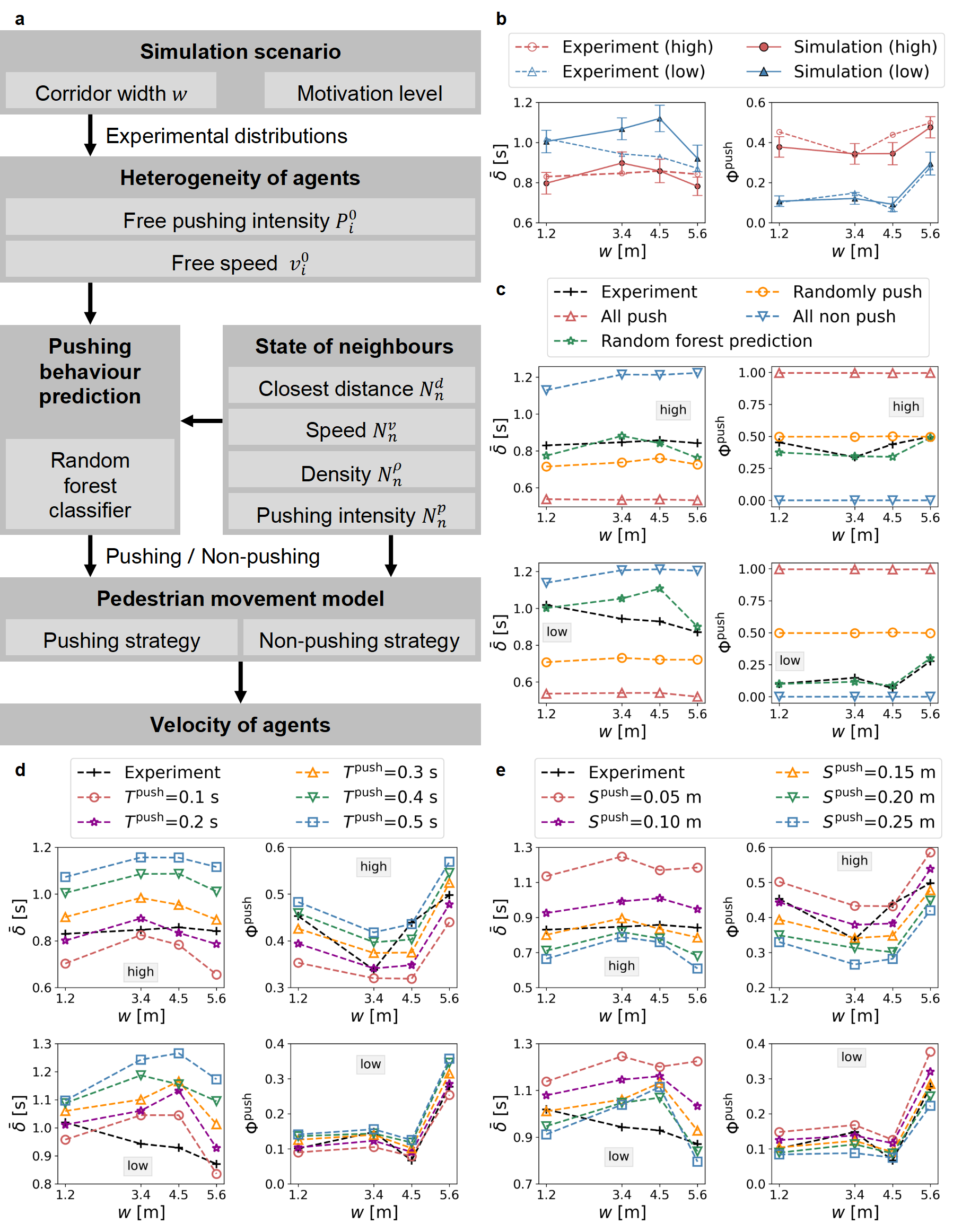}
    \caption{
    Hybrid machine learning and physics-based model.
    a, The structure of the Machine Learning Pushing Velocity Model  (ML-PVM).
    b, The proportion of pushing behaviours $\Phi^{\rm{push}}$ and the mean time lapse $\bar{\delta}$ measured from experiments and simulations in the eight scenarios. 
    c, The simulated values of $\Phi^{\rm{push}}$ and $\bar{\delta}$ when all agents engage in pushing behaviours (All push), all agents engage in non-pushing behaviours (All non push), and agents engage in pushing behaviours randomly (Randomly push).
    d, The impact of the parameter $T^{\rm{push}}$ on $\Phi^{\rm{push}}$ and $\bar{\delta}$.
    e, The impact of the parameter $S^{\rm{push}}$ on $\Phi^{\rm{push}}$ and $\bar{\delta}$.
    }
    \label{fig:ModelCali}
\end{figure}

From the previous experimental analysis, three main conclusions are drawn.
a. The distribution of pedestrian free pushing intensities is influenced by the width of the corridor and the motivation level.
b. Whether pedestrians engage in pushing behaviours can be predicted using a random forest classifier based on pedestrian free pushing intensities and vectors describing the state of neighbours. 
c. Pushing behaviours lead to a more aggressive space-utilization movement strategy that enables higher speeds by compressing the space between pedestrians.
These findings form the basis for the proposed hybrid machine learning and physics-based model, named the Machine Learning Pushing Velocity Model (ML-PVM).
The structure of the ML-PVM is illustrated in Fig.~\ref{fig:ModelCali}a.
First, the heterogeneity of agents is captured through the distributions of the free pushing intensities $P_i^0$ and the free speeds $v_i^0$.
These distributions are derived from experimental data and adapt to simulation scenarios, accounting for the width of the corridor $w$ and the motivation level.
Next, the state of neighbours is quantitatively described for each agent using the spatial discretization method previously proposed. 
Descriptive features include the closest distance between the agent and neighbours $N^d_n$, the mean effective speed of neighbours $N^v_n$, the mean density of neighbours $N^\rho_n$, and the mean pushing intensity of neighbours $N^p_n$.
Subsequently, a random forest classifier, trained with the experimental data, predicts whether an agent will engage in pushing behaviours based on its free pushing intensity and the vector describing the state of its neighbours. 
Finally, a pedestrian movement model with two distinct strategies determines the velocity of the agent, corresponding to its engagement in either pushing or non-pushing behaviours.

The pedestrian movement model adopted in the ML-PVM is based on the Generalized Collision-Free Velocity Model (GCVM)~\cite{xu2019generalized}, which is a physics-based model.
In the GCVM, for a given agent $i$, the direction of movement is first calculated,  then the speed in the direction of movement (denoted with $v_i$) is determined by
\begin{equation}
    v_i=\min\Big(v_i^0,\max\big(0,\frac{s_i}{T}\big)\Big),
\end{equation}
where $s_i$ represents the available distance of agent $i$ in its direction of movement, and $T$ is a parameter that controls the slope of the relationship between the speed and the available distance (speed-headway relationship).
A smaller $T$ allows agents to achieve higher speeds for the same amount of available distance.
To realize distinct movement strategies for pushing and non-pushing behaviours, the ML-PVM extends this speed function to
\begin{equation}
    v_i=\begin{cases}
    \min\Big(
    v_i^0,
    \max\big(
    0,
    \frac
    {s_i+S^{\rm{push}}}
    {T^{\rm{push}}}
    \big)
   \Big)
   ,& \rm{pushing}\\
   \min\Big(
    v_i^0,
    \max\big(
    0,
    \frac
    {s_i+S^{\rm{nonp}}}
    {T^{\rm{nonp}}}
    \big)
   \Big)
   ,& \rm{non\mbox{-}pushing},
    \end{cases}
\end{equation}
where parameters $S^{\rm{push}}$, $T^{\rm{push}}$, $S^{\rm{nonp}}$, and $T^{\rm{nonp}}$ govern the speed-headway relationship under each behaviour type. 
A detailed introduction to the ML-PVM is provided in Methods.  
Using a grid search method, the ML-PVM is calibrated based on two indicators: the proportion of pushing behaviours (denoted as $\Phi^{\rm{push}}$) and the mean time lapse between two consecutive agents entering the entrance (denoted as $\bar{\delta}$).
Calibration involves minimizing the Mean Absolute Error of the two indicators in the eight experimental scenarios with varying widths of the corridor and motivation levels. 
The optimal parameters determined through this process are $S^{\rm{push}}=0.15~\rm{m}$  , $T^{\rm{push}}=0.2~\rm{s}$, $S^{\rm{nonp}}=0.08~\rm{m}$, and $T^{\rm{nonp}}=0.3~\rm{s}$.
The values of additional parameters are detailed in Methods. 

Fig.~\ref{fig:ModelCali}b compares the proportion of pushing behaviours $\Phi^{\rm{push}}$ and the mean time lapse $\bar{\delta}$ measured from experiments and simulations in the eight scenarios. 
The simulated values represent the mean results of 20 simulations, each with varied random seeds to alter the initial positions of agents, as well as their free pushing intensities and free speeds.
The ML-PVM accurately reproduces the proportion of pushing behaviours in the eight scenarios with a single parameter set.
Regarding the mean time lapse, the simulation results align well with the experimental data under high-motivation scenarios. 
However, discrepancies emerge in low-motivation scenarios, particularly when the widths of the corridor are 3.4 and 4.5 meters, where the simulated exceed the experimental values.
This deviation may be attributed to the absence of cooperation between agents. 
In the scenarios where the discrepancies are most pronounced, the experimental pushing proportions are relatively low, suggesting that pedestrians rely more on cooperative strategies, such as giving their turn to others, to resolve conflicts near the entrance and achieve smoother movement. 
This cooperative dynamic has not yet been incorporated into the ML-PVM.

In addition to using the trained random forest classifier to predict whether agents engage in pushing behaviours, simulations were also conducted using three alternative approaches: All agents engage in pushing behaviours, all agents engage in non-pushing behaviours, and agents engage in pushing behaviours randomly.
Fig.~\ref{fig:ModelCali}c compares the proportion of pushing behaviours $\Phi^{\rm{push}}$ and the mean time lapse $\bar{\delta}$ in simulations conducted with these approaches to the simulations using the random forest classifier.
When all agents push, the simulated values of $\bar{\delta}$ are significantly lower than the experimental values, indicating an overly rapid flow.
In contrast, when no agents push, the values of $\bar{\delta}$ are much higher than the experimental values, reflecting a relatively slow flow.
The randomly pushing approach achieves moderate agreement in high-motivation scenarios but deviates significantly in low-motivation scenarios. 
These results indicate the necessity of accurately reproducing the proportion of pushing behaviours in simulations to achieve realistic predictions of the mean time lapse, suggesting that the crowd dynamics depend heavily on the movement strategies adopted by pedestrians.

It is important to note that the simulations in all eight scenarios are performed using a single set of parameters, although the performance could be improved by optimizing the parameters for each specific scenario.
Fig.~\ref{fig:ModelCali}d and Fig.~\ref{fig:ModelCali}e explore the impact of the parameters $T^{\rm{push}}$ and $S^{\rm{push}}$ on the proportion of pushing behaviours $\Phi^{\rm{push}}$ and the mean time lapse $\bar{\delta}$ in the eight scenarios, while other parameters remain fixed at their optimal values.
These two parameters are chosen because they are most directly related to pushing behaviours.
The analysis reveals that as $T^{\rm{push}}$ decreases and $S^{\rm{push}}$ increases, both $\bar{\delta}$ and $\Phi^{\rm{push}}$ increase consistently in all scenarios.
These trends can be explained as follows.
A smaller $T^{\rm{push}}$ and a larger $S^{\rm{push}}$ allow agents to achieve higher speeds with the same available distance in front, leading to reduced $\bar{\delta}$.
At the same time, the improved speed advantage of a single push due to a smaller $T^{\rm{push}}$ and a larger $S^{\rm{push}}$ reduces the need for multiple pushing.
Therefore, agents can reach the entrance with fewer pushing behaviours, which results in a decrease in $\Phi^{\rm{push}}$.

\subsection{Comparisons between simulations and experiments} 
\begin{figure}[htbp]
    \centering
    \includegraphics[width=\linewidth]{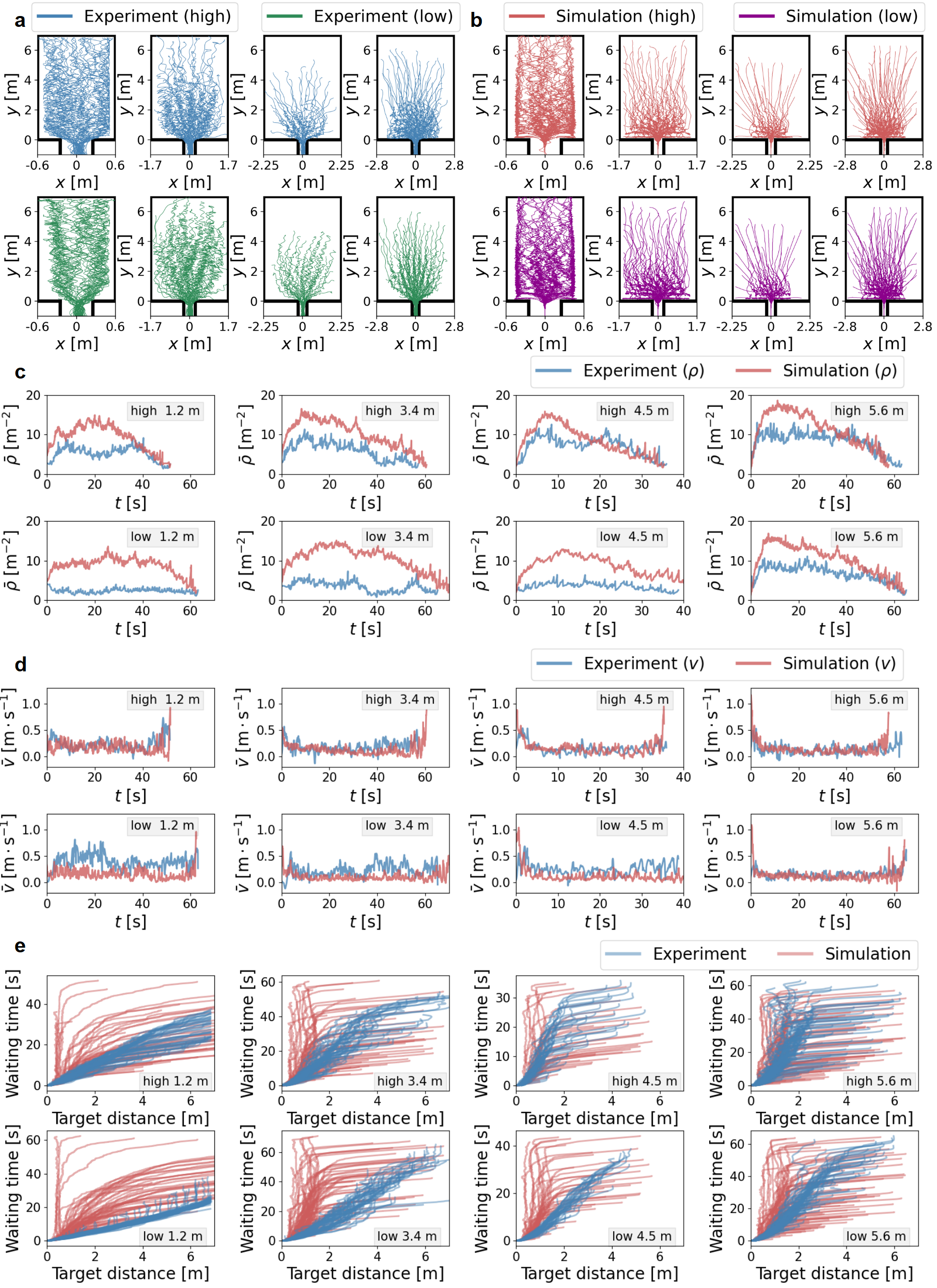}
    \caption{Comparisons between simulations and experiments.
    a, Trajectories of pedestrians in experiments.
    b, Trajectories of agents in simulations.
    c, Time series of the mean density in the measurement area.
    d, Time series of the mean effective speed in the measurement area.
    e, The relationships between the waiting time and the target distance.
    }
    \label{fig:ModelVali}
\end{figure}

To further validate the ML-PVM, simulations conducted with the ML-PVM are compared with experiments in Fig.~\ref{fig:ModelVali}, where each subfigure contains the results of eight scenarios with varying widths of corridor ($w=1.2,~3.4,~4.5,~5.6$ m) and motivation levels (high/low).
The first row shows high-motivation scenarios, while the second shows low-motivation scenarios.
From left to right, the columns correspond to scenarios with corridor widths of 1.2, 3.4, 4.5, and 5.6 meters.

The individual trajectories in the simulations (shown in Fig.~\ref{fig:ModelVali}b) are compared with those from the experiments (shown in Fig.~\ref{fig:ModelVali}a).
Two primary differences are observed.
First, while agents in the simulations are broadly distributed throughout the corridor near the $y=0$ line, experimental pedestrians show a wedge-shaped distribution near the entrance. 
This discrepancy could be attributed to the barricades used in the experiment, which include metal bases for support (Fig.~\ref{fig:expDatCol}b). 
These metal bases appear to be avoided by pedestrians, particularly in low-motivation scenarios, likely acting as obstacles absent in the simulations~\cite{rzezonka2022attempt}.
Second, the queuing phenomenon is more pronounced in the experimental trajectories, particularly under low-motivation conditions. 
One possible explanation is the lack of a movement strategy to follow others in the ML-PVM, which is critical to reproducing the phenomenon of queuing. 
To maintain simplicity, the ML-PVM incorporates only two movement strategies that correspond to pushing and non-pushing. 
All behaviours other than pushing are grouped under non-pushing, and the assigned movement strategy for non-pushing behaviours does not account for the dynamics of following others.

In addition, the mean density $\bar{\rho}$ and the mean effective speed $\bar{v}$ of pedestrians/agents within the measurement area (a rectangular area of $0.8\times0.8$ meters in front of the entrance) are analysed for both experiments and simulations.
Fig.~\ref{fig:ModelVali}c shows the time series of the mean density in the measurement area and Fig.~\ref{fig:ModelVali}d shows the time series of the mean effective speed in the measurement area.
The simulations generally exhibit higher densities compared to the experiments, and this discrepancy is more pronounced in low-motivation scenarios.  
With regard to the effective speed in the measurement area, the simulations align closely with the experimental values in high-motivation scenarios. 
However, in low-motivation scenarios, particularly when the widths of the corridor are 1.2, 3.4, and 4.5 meters, the effective speeds in the simulations are noticeably lower than those observed in experiments.
This difference arises because, in high-motivation scenarios, the high proportion of pushing behaviours dominates the crowd dynamics, enabling the proposed model to closely replicate the time series observed in the experiments. 
However, in low-motivation scenarios, where the proportion of pushing behaviours is relatively low, the crowd dynamics are mainly influenced by other behaviours, such as cooperation with others and following others, which are not fully incorporated into the ML-PVM. 
This limitation results in a greater deviation from the experimental observations. 
Nevertheless, this limitation represents an intentional trade-off, as the primary objective of the ML-PVM is to reproduce the pedestrian pushing behaviour and its influence on the crowd dynamics.

Fig.~\ref{fig:ModelVali}e further illustrates this limitation by presenting the relationships between the waiting time and the target distance for each pedestrian/agent.
Here, waiting time is defined as the remaining time until a pedestrian or an agent enters the entrance, while target distance refers to the direct distance to the position entering the entrance along a straight line~\cite{adrian2020crowds}. 
Each line represents the progression of these two values for a pedestrian or an agent.
A line more approximately parallel to the horizontal axis indicates that the pedestrian/agent is approaching the entrance at a faster rate.
In high-motivation scenarios, the differences between simulations and experiments are present but not pronounced.
In both conditions, the overall crowd dynamics are similar: pedestrians/agents initially approach the entrance rapidly but then wait near the entrance for a significant amount of time due to congestion.
In contrast, in low-motivation scenarios, the dynamics deviate more substantially.
In simulations, crowd dynamics remain similar to that observed in high-motivation scenarios, with agents initially moving quickly toward the entrance before experiencing congestion.
However, in the experiments, the pedestrians exhibit a different strategy: instead of rushing towards the entrance, they maintain a more uniform speed and gradually approach the entrance, effectively reducing the congestion near the exit.
This strategy, which is not captured in the ML-PVM, further emphasizes the significant role of various non-pushing behaviours in shaping crowd dynamics, particularly in low-motivation scenarios.

\section{Discussion}\label{sec3}
This work builds upon the assumption that the pushing intensity of a pedestrian depends on its free pushing intensity reflecting its internal pushing tendency, and the external effects from its neighbouring pedestrians.

The bottleneck experiments conducted by the research group at Forschungszentrum J\"ulich provide trajectory data for pedestrians in eight scenarios, varying by motivation levels and widths of the corridor. 
Using a four-stage behaviour categorization system, pedestrian pushing intensities are quantified at different locations and times.
The average pushing intensity of a pedestrian throughout the experiment is used to represent the free pushing intensity of the pedestrian.
The distribution of pedestrian free pushing intensity varies with the experimental scenarios, reflecting pedestrians are likely to show a stronger internal pushing tendency in the scenario with a high motivation and a wider corridor.  
In addition, pedestrian free pushing intensities significantly influence their dynamics, particularly in high-motivation scenarios.

To describe the external effects of neighbouring pedestrians, a spatial discretization method is developed that characterizes the state of neighbours using a feature vector. 
This method facilitates the building of random forest-based classifiers to predict whether a pedestrian would engage in pushing behaviours. 
Performance analysis highlights that incorporating the free pushing intensity alongside the vector of neighbour states enhances prediction accuracy (up to 0.872), compared to relying solely on the vector of neighbour states (accuracy of 0.758), which validates the initial assumption.

Further examination of pedestrian speeds and densities distributions, as well as the relationships between speeds and available distances, reveal distinct differences in movement strategies between pushing and non-pushing behaviours. 
Pedestrians who engage in pushing adopt an aggressive space-utilization movement strategy, compressing the distance to neighbours to achieve higher speeds.

Based on these findings, the hybrid machine learning and physics-based model named the Machine Learning Pushing Velocity Model (ML-PVM) is introduced, which incorporates three features absent in other microscopic models for pedestrian dynamics.
a. Heterogeneity in internal pushing tendencies: The free pushing intensities of agents are sampled from experimental distributions linked to motivation levels and widths of the corridor.
b. Prediction of pushing behaviours: A random forest classifier is incorporated to predict whether an agent engages in pushing behaviour based on its free pushing intensity and the state of its neighbours.
c. Incorporation of multi-strategies: The velocities of agents are calculated using two distinct rules, reflecting the movement strategies of pushing and non-pushing behaviours.
In addition to these innovations, the ML-PVM integrates the rational part of velocity-based models (speed-headway relationships) and force-based models (contact forces). 
The ML-PVM accurately reproduces experimental observations, including the proportion of pushing behaviours and the mean time lapse between two consecutive agents entering the entrance, realizing by adopting a set of parameters in eight scenarios with varying widths of the corridor and motivation levels. 
The ML-PVM also demonstrates flexibility in optimizing performance through scenario-specific parameter adjustments.

Despite these advances, limitations remain. 
The model effectively captures crowd dynamics in high-motivation scenarios, where pushing behaviour predominates, but discrepancies arise in low-motivation scenarios, where non-pushing behaviours such as cooperation and following are more dominant. 
Since the manual categorization by Üsten et al. \cite{usten2022pushing} focused solely on the presence or absence of pushing behaviours, other behaviours were not incorporated into the ML-PVM as well. 
Future research will focus on designing experiments and observation methods to better capture these behaviours and incorporate the corresponding movement strategies into the ML-PVM. 
Furthermore, while the random forest classifier achieves a prediction accuracy of 0.872, exploring more advanced classification algorithms may further enhance performance.
This will be an important issue for future development.
The introduction of pushing behaviours into the ML-PVM has expanded the model parameter space, requiring the calibration of 11 parameters.
Introducing additional behaviours like cooperation and following is expected to increase this complexity further, posing challenges for parameter optimization. 
To address this, future work will explore parameter fitting methodologies on the basis of reinforcement learning techniques to improve efficiency.

In summary, this study provides a significant step forward in understanding and modelling pedestrian pushing behaviours, offering valuable information on the underlying mechanisms and movement strategies. 
The proposed ML-PVM successfully captures pedestrian dynamics in high-motivation scenarios and, with future improvements to overcome its current limitations, holds promise for achieving more accurate simulations and broader applicability across diverse scenarios and domains.

\section{Methods}\label{sec4} 
\subsection{Experimental participants and instructions}
Participants for the experiments were recruited from the University of Wuppertal and rewarded with a dining hall voucher valued at approximately 5 euros. 
Each participant took part in only one experimental setup, which included two runs: one with high motivation and another with low motivation, both conducted for the same corridor width.
For experiments with corridor widths of 1.2, 3.4, 4.5, and 5.6 meters, the number of participants was 63, 67, 42, and 75, respectively.
The instruction for the high-motivation experiments was: ``Imagine you are on your way to a concert by your favourite artist. 
You know that at the back you can hardly see anything at all or only the video screen. 
You absolutely want to be standing next to the stage and therefore want to access the concert as fast as possible.''
The instruction for the low-motivation experiments was: ``Imagine again that you are on your way to a concert by your favourite artist. 
This time you know that everyone will have a good view. 
However, you would like to access the concert quickly''
~\cite{adrian2020crowds}
All instructions were read to participants in German before they moved.
It should be noted that the original data set comprises 14 experimental data sets.
For this study, 8 were selected where pedestrians were instructed to form a crowded configuration, while the remaining 6, in which pedestrians were instructed to form queues, were excluded.

\subsection{Definition of the ML-PVM}
The hybrid model proposed in this work is named the Machine Learning Pushing Velocity Model, abbreviated as ML-PVM.
Agents are represented by circular disks with a constant radius $r$.
The position and velocity of the agent $i$ are defined as $\vec{x}_i$ and $\vec{v}_i$, respectively.
The relationship is $\vec{v}_i=\dot{\vec{x}}_i=\vec{e}_i \cdot v$, where $\vec{e}_i$ is the direction of movement and $v$ is the speed.
The velocity and position of agents are updated according to the following rules at each time step of the simulation.

a. Predicting whether agents will engage in pushing behaviours:
Influenced by pedestrian internal pushing tendencies and external effects from neighbours, pedestrians could transfer between pushing and non-pushing behaviours.
The ML-PVM introduces this transformation by predicting the behaviour of agent $i$ as pushing or non-pushing using a binary random forest classifier trained on experimental data, denoted as $C^{\rm{RF}}$.
Since the classifier $C^{\rm{RF}}$ requires the pushing intensity of neighbouring agents as input, the behaviours of agents are mapped to corresponding pushing intensity values.
In the previous analysis, the ordinal numbers assigned to behaviours in the four-stage category system~\cite{usten2022pushing} are used to represent the pushing intensities of various behaviours.
The pushing intensities of the category falling behind, just walking, mild pushing, and strong pushing are 1, 2, 3, and 4 respectively. 
However, as the classifier performs binary classification, 
falling behind and just walking are grouped as non-pushing behaviours, mild and strong pushing are grouped as pushing behaviours.
Considering the training data set contains significantly fewer samples with pushing intensities of 1 and 4  compared to 2 and 3 , the pushing intensity of agent $i$, denoted as $P_i$, is set as
\begin{equation}
    P_i=
    \begin{cases}
        3, & C^{\rm{RF}}(P^0_i,N_i)=\rm{Pushing}\\
        2, & C^{\rm{RF}}(P^0_i,N_i)=\rm{Non\mbox{-}pushing}.
    \end{cases}
\end{equation}
The input features of the classifier $C^{\rm{RF}}$ include $P^0_i$, the free pushing intensity of agent $i$, and $N_i=[N^d_{i,1}, N^v_{i,1}, N^\rho_{i,1}, N^p_{i,1}, \cdots, N^d_{i,n}, N^v_{i,n}, N^\rho_{i,n}, N^p_{i,n}]$, a vector that quantitatively describes the state of neighbours using the space discretization method shown in 
Fig.~\ref{fig:neigEffe}a.
In the vector $N_i$, $N^d_{i,n}$ is the closest distance between the agent $i$ and the neighbours located in the region $n$, $N^v_{i,n}$, $N^\rho_{i,n}$, and $N^p_{i,n}$ are the mean speed, the mean density, and the mean pushing intensity of agent $i$'s neighbours that are located in the region $n$.

b. Calculating the preferable velocities of agents according to their behaviours:
To reach the target efficiently and avoid conflicts with neighbours, pedestrians could actively adjust their current velocities to their preferred velocities. 
The adjustment strategies should be different under pushing and non-pushing behaviours, which is reflected in the ML-PVM by adopting two sets of calculation parameters.
The calculation process of agent $i$'s preferred velocity is as follows.

First, the desired direction of agent $i$, denoted as $\vec{e}^{~0}_i$, is defined as a unit vector that points from its position $\vec{x}_i$ to its target.  
The determination of the target could follow various tactical strategies.
For the bottleneck structure scenario studied in this paper, the target is defined as the midpoint of the entrance.

\begin{figure}[htbp]
    \centering
    \includegraphics[width=\linewidth]{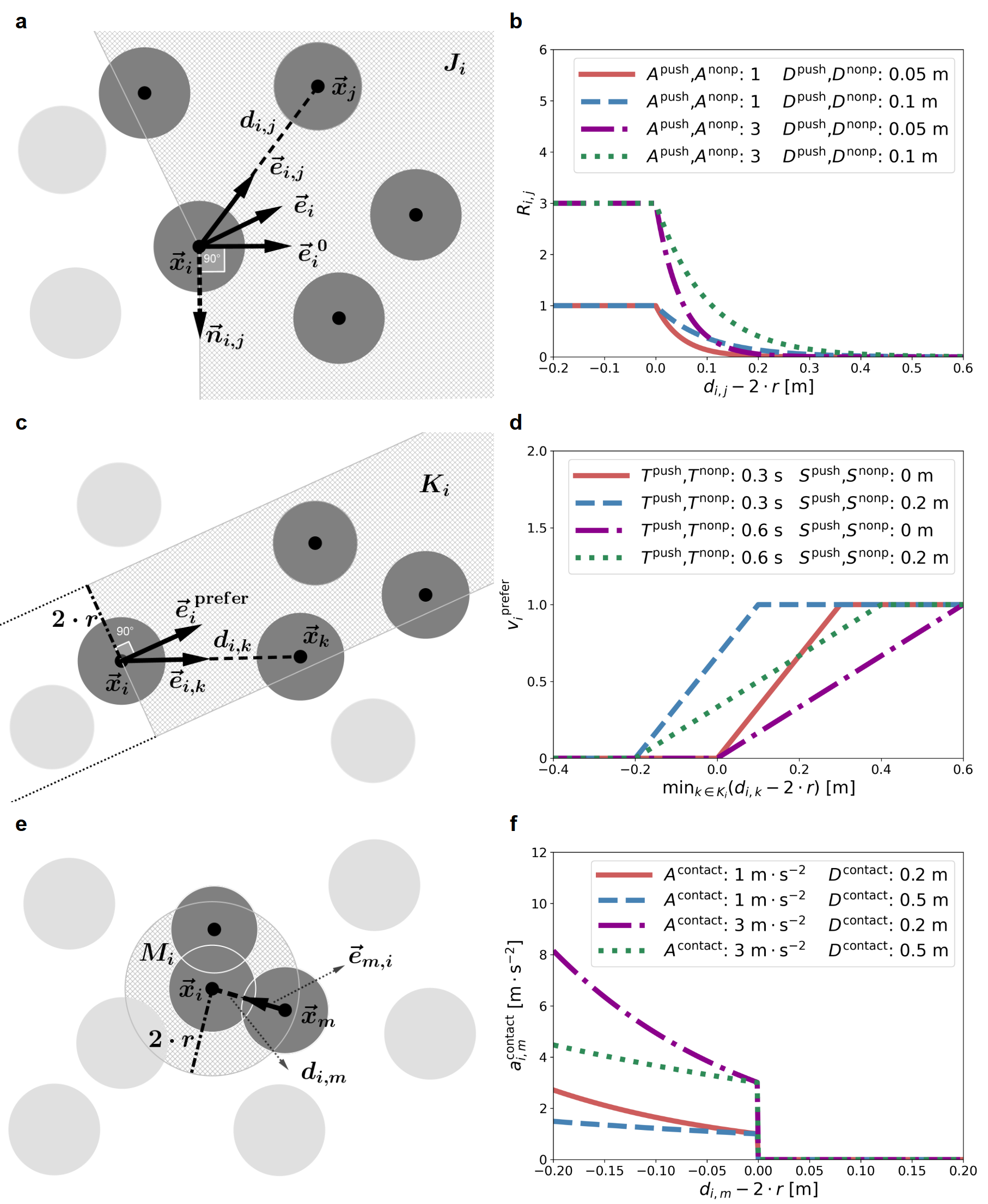}
    \caption{Definition of the ML-PVM.
    a, The calculation method for the impact from neighbours on agent $i$'s direction of movement.
    b, The relationships between $R_{i,j}$ and $d_{i,j}-2 \cdot r$.
    c, The calculation method for agent $i$'s preferred speed in its preferred direction of movement.
    d, The relationships between $v_i^{~\rm{prefer}}$ and $\min_{k\in K_i}(d_{i,k}-2 \cdot r)$.
    e, The calculation method for agent $i$'s forced acceleration caused by the contacts with neighbours.
    f, The relationships between $a_{i,m}^{~\rm{contact}}$ and $d_{i,m}-2 \cdot r$. 
    }
    \label{fig:ModelDefi}
\end{figure}

Second, the impact of neighbours on the direction of movement is calculated.
Fig.~\ref{fig:ModelDefi}a describes the calculation method.
The set of neighbours that have an impact on agent $i$'s direction of movement is defined as
\begin{equation}
     J_i=\left\{j,~\vec{e}_i \cdot \vec{e}_{i,j}>0~\text{or}~  \vec{e}_i^{~0} \cdot \vec{e}_{i,j}>0\right\},
\end{equation}
where $\vec{e}_i$ is agent $i$'s direction of movement and $\vec{e}_{i,j}$ is the unit vector pointing from the position of agent $i$ to agent $j$. 
The direction of the impact from agent $j$ is defined as 
\begin{equation}
    \vec{n}_{i,j}=
    -{\rm{sgn}}
    (\vec{e}_{i,j} \cdot \vec{e}_i^{~0\bot}) \cdot\vec{e}_i^{~0\bot},
\end{equation}
and the strength of this impact is defined as
\begin{equation}
    R_{i,j}=\begin{cases}
        A^{\rm{push}} ,& P_i=3{~\rm{and}~} d_{i,j}<2 \cdot r\\
        A^{\rm{push}} \cdot \exp(\frac{2 \cdot r-d_{i,j}}{D^{\rm{push}}}) ,& P_i=3{~\rm{and}~} d_{i,j}\geq2 \cdot r\\
        A^{\rm{nonp}} ,& P_i=2{~\rm{and}~} d_{i,j}<2 \cdot r\\
        A^{\rm{nonp}} \cdot \exp(\frac{2 \cdot r-d_{i,j}}{D^{\rm{nonp}}})
        ,& P_i=2{~\rm{and}~} d_{i,j}\geq2 \cdot r,
    \end{cases}
\end{equation}
where $d_{i,j}=\lVert\vec{x}_k-\vec{x}_i\rVert$ is the distance between agent $i$ and $j$, $A^{\rm{push}}$, $D^{\rm{push}}$, $A^{\rm{nonp}}$, and $D^{\rm{nonp}}$ are parameters used to calibrate the range of the impact from neighbours.
Fig.~\ref{fig:ModelDefi}b shows the relationships between $R_{i,j}$ and $d_{i,j}-2 \cdot r$ with various values of $A^{\rm{push}}$(or $A^{\rm{nonp}}$) and $D^{\rm{push}}$ (or $D^{\rm{nonp}}$).

Third, agent $i$'s preferred direction of movement is calculated as 
\begin{equation}
    \vec{e}_i^{~\rm{prefer}} = \vec{e}_{i}+\frac{ u \cdot (\vec{e}_i^{~0}+\sum_{j \in J_i}{R_{i,j} \cdot \vec{n}_{i,j}})-\vec{e}_{i}}{\tau}\cdot \delta t,
\end{equation}
where $u$ is a normalization constant such that $|| u \cdot (\vec{e}_i^{~0}+\sum_{j \in J_i}{R_{i,j} \cdot \vec{n}_{i,j}})||=1$, $\tau$ is a relaxation parameter adjusting the rate of agents' turning process, and $\delta t$ is the time step size of simulations.

Fourth, agent $i$'s preferred speed in its preferred direction of movement is calculated.
Fig.~\ref{fig:ModelDefi}c describes the calculation method.
The set of neighbours that would affect the speed at which agent $i$ moves in its preferred direction of movement is defined as
\begin{equation}
    K_i=\left\{k,~\vec{e}_i^{~\rm{prefer}} \cdot \vec{e}_{i,k} \ge 0\ \text{and}\ \left|\vec{e}^{~\rm{prefer}\bot} \cdot \vec{e}_{i,k}\right| \leq \frac{2\cdot r}{d_{i,k}}\right\}.
\end{equation}
Considering pedestrians who engage in pushing behaviours prefer to intrude into the space occupied by neighbours than those who take non-pushing behaviours, agent $i$'s preferred speed in its preferred direction of movement is calculated as 
\begin{equation}
    v_i^{~\rm{prefer}}=\begin{cases}
    \min\Big(
    v_i^0,
    \max\big(
    0,
    \frac
    {\min_{k\in K_i}(d_{i,k}-2 \cdot r)+S^{\rm{push}}}
    {T^{\rm{push}}}
    \big)
   \Big)
   ,& P_i=3\\
   \min\Big(
    v_i^0,
    \max\big(
    0,
    \frac
    {\min_{k\in K_i}(d_{i,k}-2 \cdot r)+S^{\rm{nonp}}}
    {T^{\rm{nonp}}}
    \big)
   \Big)
   ,& P_i=2,
    \end{cases}
\end{equation}
where $v_i^0$ is the free speed of agent $i$, $\min_{k\in K_i}(d_{i,k}-2 \cdot r)$ is the maximum distance that agent $i$ can move in its preferred direction of movement without overlapping other agents,  $S^{\rm{push}}$, $T^{\rm{push}}$, $S^{\rm{nonp}}$, and $T^{\rm{nonp}}$ are parameters used to calibrate the speed-headway relationship.
Fig.~\ref{fig:ModelDefi}d gives the relationships between $v_i^{~\rm{prefer}}$ and $\min_{k\in K_i}(d_{i,k}-2 \cdot r)$ with various values of $S^{\rm{push}}$(or $S^{\rm{nonp}}$) and $T^{\rm{push}}$ (or $T^{\rm{nonp}}$).

c. Integrating the preferable velocities of agents and the forced acceleration caused by contacts with neighbours:
Pushing behaviours lead to contact between pedestrians, which prevents pedestrians from moving at their preferred velocities. 
The forced acceleration caused by the contacts between agents is calculated as follows.
Fig.~\ref{fig:ModelDefi}e describes the calculation method.
The set of neighbours that have contact with agent $i$ is defined as 
\begin{equation}
    M_i=\left\{m,d_{i,m}<2 \cdot r\right\}.
\end{equation}
The direction of agent $i$'s acceleration due to the contact with agent $k$ is set as $\vec{e}_{m,i}$, which is the unit vector pointing from the position of agent $m$ to agent $i$.
The strength of this acceleration is defined as
\begin{equation}
    a^{~\rm{contact}}_{i,m}=A^{\rm{contact}} \cdot \exp(\frac{2 \cdot r-d_{i,m}}{D^{\rm{contact}}}),
\end{equation}
where $A^{\rm{contact}}$, and $D^{\rm{contact}}$ are parameters used to calibrate the range of the acceleration caused by contacts with neighbours.
Fig.~\ref{fig:ModelDefi}f shows the relationships between $a_{i,m}^{~\rm{contact}}$ and $d_{i,m}-2 \cdot r$ with various values of $A^{\rm{contact}}$ and $D^{\rm{contact}}$.
Finally, the velocity of agent $i$ is updated as 
\begin{equation}
    \vec{V}_i=\vec{e}_i^{~\rm{prefer}} \cdot v_i^{\rm{prefer}}+\sum_{m \in M_i} \vec{e}_{m,i} \cdot a^{~\rm{contact}}_{i,m} \cdot \delta t.
\end{equation}

\subsection{Simulation parameters}
In simulations, the free pushing intensities $P_i^0$ and the free speeds $v_i^0$ of agents are set to follow the experimental distributions.
The distribution of $P_i^0$ for each simulated scenario is described by the bimodal distribution function as defined in Eq.~\ref{equ:bimodalDist}.
Table~\ref{tab:ModelPara}a lists the parameters for the eight simulated scenarios.
In this table, the values in the first column correspond to $A_1$, $\mu_1$, $\sigma_1$ from top to bottom, while the values in the second column correspond to $A_2$, $\mu_2$, $\sigma_2$ from top to bottom.
The distribution of $v_i^0$ for each simulated scenario is based on the maximum speeds of each pedestrian observed in the corresponding experiment, modeled using a Gaussian distribution function.
Table~\ref{tab:ModelPara}b presents the parameters for the eight simulated scenarios, where the first value represents the mean and the second value denotes the standard deviation.
For the random forest classifier, to simplify the implementation and improve the simulation efficiency, the anticipation time $t^{\rm{anti}}$ and the number of discrete regions $n$ are fixed at 0 seconds and 2, respectively.
The simulation time step size $\delta t$ is set to 0.04 seconds, which corresponds to 25 frames per second to match the experimental data. 
The radius of the agent was set to 0.18 meters~\cite{xu2021anticipation}.
Other parameters are optimized using a grid search method. 
Parameters were grouped according to their importance, allowing gradual refinement to identify optimal values. 
This stepwise optimization ensured efficient calibration while maintaining model accuracy.
The final set of optimal parameters, consistent across the eight simulated scenarios, is provided in Table~\ref{tab:ModelPara}c.

\begin{table}[htbp]
    \centering
    \caption{Simulation parameters.
    a, The parameters of agents' free pushing intensity $P_i^0$ distribution function.
    b, The parameters of agents' free speed $v_i^0$ distribution function.
    c, The optimal simulation parameters.}
    \begin{tabular}{c}
         \begin{minipage}[b]{\linewidth}\raisebox{-0.5\height}{\includegraphics[width=\linewidth]{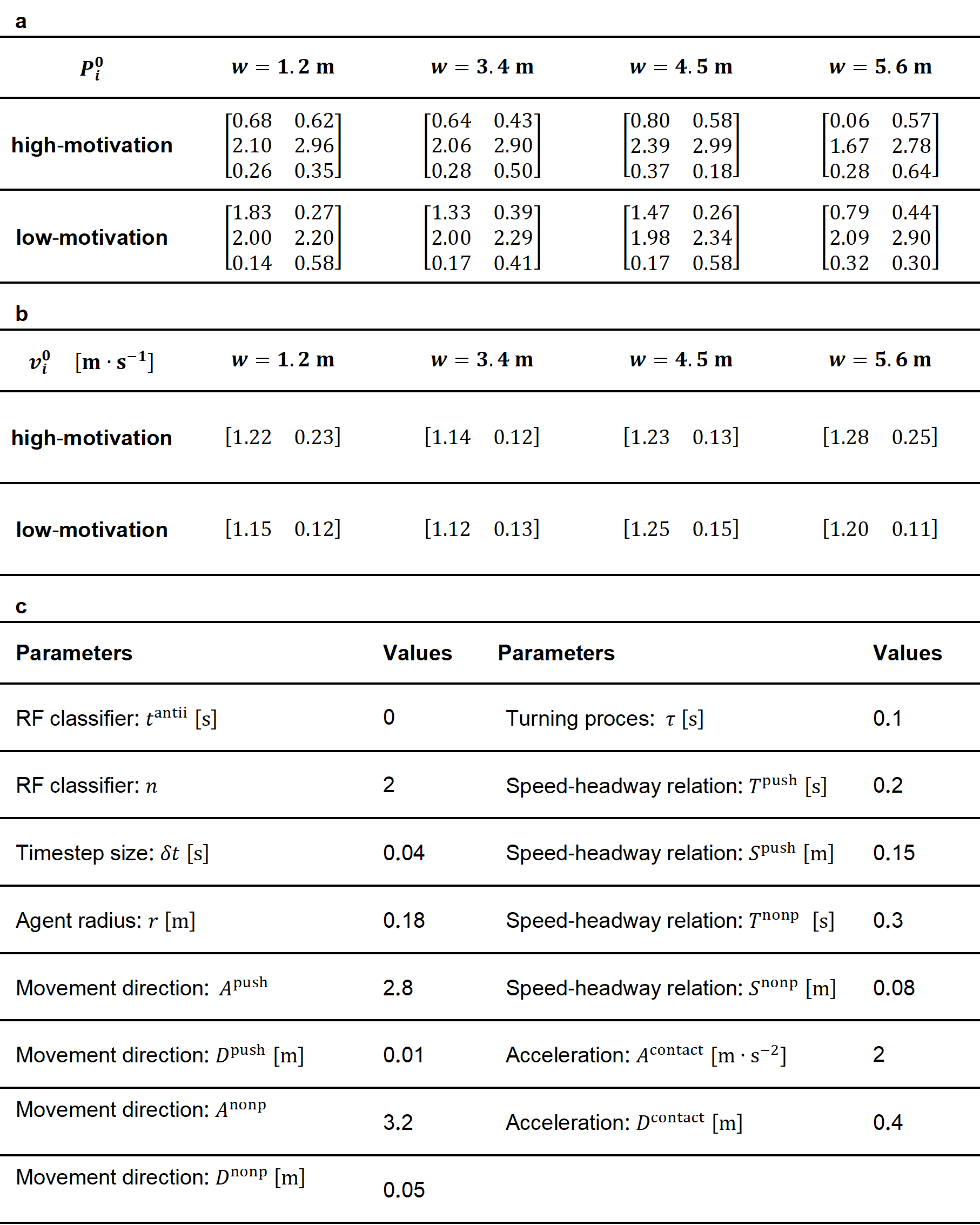}}
         \end{minipage}
    \end{tabular}
    \label{tab:ModelPara}
\end{table}

\backmatter

\bibliography{sn-bibliography}

\begin{thebibliography}{10}
\expandafter\ifx\csname url\endcsname\relax
  \def\url#1{\burl{#1}}\fi
\expandafter\ifx\csname urlprefix\endcsname\relax\def\urlprefix{URL }\fi
\providecommand{\bibinfo}[2]{#2}
\providecommand{\eprint}[2][]{\url{#2}}
\providecommand{\doi}[1]{\url{https://doi.org/#1}}
\bibcommenthead

\bibitem{sieben2023inside}
\bibinfo{author}{Sieben, A.} \& \bibinfo{author}{Seyfried, A.}
\newblock \bibinfo{title}{Inside a life-threatening crowd: Analysis of the love parade disaster from the perspective of eyewitnesses}.
\newblock \emph{\bibinfo{journal}{Safety science}} \textbf{\bibinfo{volume}{166}}, \bibinfo{pages}{106229} (\bibinfo{year}{2023}).

\bibitem{helbing2007dynamics}
\bibinfo{author}{Helbing, D.}, \bibinfo{author}{Johansson, A.} \& \bibinfo{author}{Al-Abideen, H.~Z.}
\newblock \bibinfo{title}{Dynamics of crowd disasters: An empirical study}.
\newblock \emph{\bibinfo{journal}{Physical Review E—Statistical, Nonlinear, and Soft Matter Physics}} \textbf{\bibinfo{volume}{75}}, \bibinfo{pages}{046109} (\bibinfo{year}{2007}).

\bibitem{zhen2008analysis}
\bibinfo{author}{Zhen, W.}, \bibinfo{author}{Mao, L.} \& \bibinfo{author}{Yuan, Z.}
\newblock \bibinfo{title}{Analysis of trample disaster and a case study--mihong bridge fatality in china in 2004}.
\newblock \emph{\bibinfo{journal}{Safety Science}} \textbf{\bibinfo{volume}{46}}, \bibinfo{pages}{1255--1270} (\bibinfo{year}{2008}).

\bibitem{illiyas2013human}
\bibinfo{author}{Illiyas, F.~T.}, \bibinfo{author}{Mani, S.~K.}, \bibinfo{author}{Pradeepkumar, A.} \& \bibinfo{author}{Mohan, K.}
\newblock \bibinfo{title}{Human stampedes during religious festivals: A comparative review of mass gathering emergencies in india}.
\newblock \emph{\bibinfo{journal}{International Journal of Disaster Risk Reduction}} \textbf{\bibinfo{volume}{5}}, \bibinfo{pages}{10--18} (\bibinfo{year}{2013}).

\bibitem{feliciani2023trends}
\bibinfo{author}{Feliciani, C.}, \bibinfo{author}{Corbetta, A.}, \bibinfo{author}{Haghani, M.} \& \bibinfo{author}{Nishinari, K.}
\newblock \bibinfo{title}{Trends in crowd accidents based on an analysis of press reports}.
\newblock \emph{\bibinfo{journal}{Safety science}} \textbf{\bibinfo{volume}{164}}, \bibinfo{pages}{106174} (\bibinfo{year}{2023}).

\bibitem{adrian2020crowds}
\bibinfo{author}{Adrian, J.}, \bibinfo{author}{Seyfried, A.} \& \bibinfo{author}{Sieben, A.}
\newblock \bibinfo{title}{Crowds in front of bottlenecks at entrances from the perspective of physics and social psychology}.
\newblock \emph{\bibinfo{journal}{Journal of the Royal Society Interface}} \textbf{\bibinfo{volume}{17}}, \bibinfo{pages}{20190871} (\bibinfo{year}{2020}).

\bibitem{lugering2023psychological}
\bibinfo{author}{L{\"u}gering, H.}, \bibinfo{author}{Alia, A.} \& \bibinfo{author}{Sieben, A.}
\newblock \bibinfo{title}{Psychological pushing propagation in crowds—does the observation of pushing behavior promote further intentional pushing?}
\newblock \emph{\bibinfo{journal}{Frontiers in Social Psychology}} \textbf{\bibinfo{volume}{1}}, \bibinfo{pages}{1263953} (\bibinfo{year}{2023}).

\bibitem{garcimartin2016flow}
\bibinfo{author}{Garcimart{\'\i}n, A.}, \bibinfo{author}{Parisi, D.~R.}, \bibinfo{author}{Pastor, J.~M.}, \bibinfo{author}{Mart{\'\i}n-G{\'o}mez, C.} \& \bibinfo{author}{Zuriguel, I.}
\newblock \bibinfo{title}{Flow of pedestrians through narrow doors with different competitiveness}.
\newblock \emph{\bibinfo{journal}{Journal of Statistical Mechanics: Theory and Experiment}} \textbf{\bibinfo{volume}{2016}}, \bibinfo{pages}{043402} (\bibinfo{year}{2016}).

\bibitem{murakami2021mutual}
\bibinfo{author}{Murakami, H.}, \bibinfo{author}{Feliciani, C.}, \bibinfo{author}{Nishiyama, Y.} \& \bibinfo{author}{Nishinari, K.}
\newblock \bibinfo{title}{Mutual anticipation can contribute to self-organization in human crowds}.
\newblock \emph{\bibinfo{journal}{Science Advances}} \textbf{\bibinfo{volume}{7}}, \bibinfo{pages}{eabe7758} (\bibinfo{year}{2021}).

\bibitem{xiao2019investigation}
\bibinfo{author}{Xiao, Y.} \emph{et~al.}
\newblock \bibinfo{title}{Investigation of pedestrian dynamics in circle antipode experiments: Analysis and model evaluation with macroscopic indexes}.
\newblock \emph{\bibinfo{journal}{Transportation Research Part C: Emerging Technologies}} \textbf{\bibinfo{volume}{103}}, \bibinfo{pages}{174--193} (\bibinfo{year}{2019}).

\bibitem{ma2021spontaneous}
\bibinfo{author}{Ma, Y.}, \bibinfo{author}{Lee, E. W.~M.}, \bibinfo{author}{Shi, M.} \& \bibinfo{author}{Yuen, R. K.~K.}
\newblock \bibinfo{title}{Spontaneous synchronization of motion in pedestrian crowds of different densities}.
\newblock \emph{\bibinfo{journal}{Nature human behaviour}} \textbf{\bibinfo{volume}{5}}, \bibinfo{pages}{447--457} (\bibinfo{year}{2021}).

\bibitem{seyfried2009new}
\bibinfo{author}{Seyfried, A.} \emph{et~al.}
\newblock \bibinfo{title}{New insights into pedestrian flow through bottlenecks}.
\newblock \emph{\bibinfo{journal}{Transportation Science}} \textbf{\bibinfo{volume}{43}}, \bibinfo{pages}{395--406} (\bibinfo{year}{2009}).

\bibitem{zuriguel2014clogging}
\bibinfo{author}{Zuriguel, I.} \emph{et~al.}
\newblock \bibinfo{title}{Clogging transition of many-particle systems flowing through bottlenecks}.
\newblock \emph{\bibinfo{journal}{Scientific reports}} \textbf{\bibinfo{volume}{4}}, \bibinfo{pages}{7324} (\bibinfo{year}{2014}).

\bibitem{helbing2005self}
\bibinfo{author}{Helbing, D.}, \bibinfo{author}{Buzna, L.}, \bibinfo{author}{Johansson, A.} \& \bibinfo{author}{Werner, T.}
\newblock \bibinfo{title}{Self-organized pedestrian crowd dynamics: Experiments, simulations, and design solutions}.
\newblock \emph{\bibinfo{journal}{Transportation science}} \textbf{\bibinfo{volume}{39}}, \bibinfo{pages}{1--24} (\bibinfo{year}{2005}).

\bibitem{haghani2019push}
\bibinfo{author}{Haghani, M.}, \bibinfo{author}{Sarvi, M.} \& \bibinfo{author}{Shahhoseini, Z.}
\newblock \bibinfo{title}{When ‘push’does not come to ‘shove’: Revisiting ‘faster is slower’in collective egress of human crowds}.
\newblock \emph{\bibinfo{journal}{Transportation research part A: policy and practice}} \textbf{\bibinfo{volume}{122}}, \bibinfo{pages}{51--69} (\bibinfo{year}{2019}).

\bibitem{usten2022pushing}
\bibinfo{author}{{\"U}sten, E.}, \bibinfo{author}{L{\"u}gering, H.} \& \bibinfo{author}{Sieben, A.}
\newblock \bibinfo{title}{Pushing and non-pushing forward motion in crowds: A systematic psychological observation method for rating individual behavior in pedestrian dynamics}.
\newblock \emph{\bibinfo{journal}{Collective dynamics}} \textbf{\bibinfo{volume}{7}}, \bibinfo{pages}{1--16} (\bibinfo{year}{2022}).

\bibitem{alia2022hybrid}
\bibinfo{author}{Alia, A.}, \bibinfo{author}{Maree, M.} \& \bibinfo{author}{Chraibi, M.}
\newblock \bibinfo{title}{A hybrid deep learning and visualization framework for pushing behavior detection in pedestrian dynamics}.
\newblock \emph{\bibinfo{journal}{Sensors}} \textbf{\bibinfo{volume}{22}}, \bibinfo{pages}{4040} (\bibinfo{year}{2022}).

\bibitem{alia2023cloud}
\bibinfo{author}{Alia, A.}, \bibinfo{author}{Maree, M.}, \bibinfo{author}{Chraibi, M.}, \bibinfo{author}{Toma, A.} \& \bibinfo{author}{Seyfried, A.}
\newblock \bibinfo{title}{A cloud-based deep learning framework for early detection of pushing at crowded event entrances}.
\newblock \emph{\bibinfo{journal}{IEEE access}} \textbf{\bibinfo{volume}{11}}, \bibinfo{pages}{45936--45949} (\bibinfo{year}{2023}).

\bibitem{alia2024novel}
\bibinfo{author}{Alia, A.}, \bibinfo{author}{Maree, M.}, \bibinfo{author}{Chraibi, M.} \& \bibinfo{author}{Seyfried, A.}
\newblock \bibinfo{title}{A novel voronoi-based convolutional neural network framework for pushing person detection in crowd videos}.
\newblock \emph{\bibinfo{journal}{Complex \& Intelligent Systems}} \bibinfo{pages}{1--27} (\bibinfo{year}{2024}).

\bibitem{xie2022experiment}
\bibinfo{author}{Xie, C.-Z.}, \bibinfo{author}{Tang, T.-Q.}, \bibinfo{author}{Zhang, B.-T.} \& \bibinfo{author}{Xiang, H.-J.}
\newblock \bibinfo{title}{Experiment, model, and simulation of the pedestrian flow around a training school classroom during the after-class period}.
\newblock \emph{\bibinfo{journal}{Simulation}} \textbf{\bibinfo{volume}{98}}, \bibinfo{pages}{63--82} (\bibinfo{year}{2022}).

\bibitem{xu2021anticipation}
\bibinfo{author}{Xu, Q.}, \bibinfo{author}{Chraibi, M.} \& \bibinfo{author}{Seyfried, A.}
\newblock \bibinfo{title}{Anticipation in a velocity-based model for pedestrian dynamics}.
\newblock \emph{\bibinfo{journal}{Transportation research part C: emerging technologies}} \textbf{\bibinfo{volume}{133}}, \bibinfo{pages}{103464} (\bibinfo{year}{2021}).

\bibitem{xu2024analysis}
\bibinfo{author}{Xu, Q.}, \bibinfo{author}{Yuan, Z.}, \bibinfo{author}{Guo, R.}, \bibinfo{author}{He, B.} \& \bibinfo{author}{Chraibi, M.}
\newblock \bibinfo{title}{Analysis and modeling of detours in pedestrian operational navigation}.
\newblock \emph{\bibinfo{journal}{Transportation Research Part C: Emerging Technologies}} \textbf{\bibinfo{volume}{162}}, \bibinfo{pages}{104584} (\bibinfo{year}{2024}).

\bibitem{wu2023force}
\bibinfo{author}{Wu, W.}, \bibinfo{author}{Yi, W.}, \bibinfo{author}{Wang, X.} \& \bibinfo{author}{Zheng, X.}
\newblock \bibinfo{title}{A force-based model for adaptively controlling the spatial configuration of pedestrian subgroups at non-extreme densities}.
\newblock \emph{\bibinfo{journal}{Transportation research part C: emerging technologies}} \textbf{\bibinfo{volume}{152}}, \bibinfo{pages}{104154} (\bibinfo{year}{2023}).

\bibitem{yates2015incorporating}
\bibinfo{author}{Yates, C.~A.}, \bibinfo{author}{Parker, A.} \& \bibinfo{author}{Baker, R.~E.}
\newblock \bibinfo{title}{Incorporating pushing in exclusion-process models of cell migration}.
\newblock \emph{\bibinfo{journal}{Physical Review E}} \textbf{\bibinfo{volume}{91}}, \bibinfo{pages}{052711} (\bibinfo{year}{2015}).

\bibitem{helbing2000simulating}
\bibinfo{author}{Helbing, D.}, \bibinfo{author}{Farkas, I.} \& \bibinfo{author}{Vicsek, T.}
\newblock \bibinfo{title}{Simulating dynamical features of escape panic}.
\newblock \emph{\bibinfo{journal}{Nature}} \textbf{\bibinfo{volume}{407}}, \bibinfo{pages}{487--490} (\bibinfo{year}{2000}).

\bibitem{breiman2001random}
\bibinfo{author}{Breiman, L.}
\newblock \bibinfo{title}{Random forests}.
\newblock \emph{\bibinfo{journal}{Machine learning}} \textbf{\bibinfo{volume}{45}}, \bibinfo{pages}{5--32} (\bibinfo{year}{2001}).

\bibitem{boltes2013collecting}
\bibinfo{author}{Boltes, M.} \& \bibinfo{author}{Seyfried, A.}
\newblock \bibinfo{title}{Collecting pedestrian trajectories}.
\newblock \emph{\bibinfo{journal}{Neurocomputing}} \textbf{\bibinfo{volume}{100}}, \bibinfo{pages}{127--133} (\bibinfo{year}{2013}).

\bibitem{steffen2010methods}
\bibinfo{author}{Steffen, B.} \& \bibinfo{author}{Seyfried, A.}
\newblock \bibinfo{title}{Methods for measuring pedestrian density, flow, speed and direction with minimal scatter}.
\newblock \emph{\bibinfo{journal}{Physica A: Statistical mechanics and its applications}} \textbf{\bibinfo{volume}{389}}, \bibinfo{pages}{1902--1910} (\bibinfo{year}{2010}).

\bibitem{devries2021using}
\bibinfo{author}{DeVries, Z.} \emph{et~al.}
\newblock \bibinfo{title}{Using a national surgical database to predict complications following posterior lumbar surgery and comparing the area under the curve and f1-score for the assessment of prognostic capability}.
\newblock \emph{\bibinfo{journal}{The spine journal}} \textbf{\bibinfo{volume}{21}}, \bibinfo{pages}{1135--1142} (\bibinfo{year}{2021}).

\bibitem{xu2019generalized}
\bibinfo{author}{Xu, Q.}, \bibinfo{author}{Chraibi, M.}, \bibinfo{author}{Tordeux, A.} \& \bibinfo{author}{Zhang, J.}
\newblock \bibinfo{title}{Generalized collision-free velocity model for pedestrian dynamics}.
\newblock \emph{\bibinfo{journal}{Physica A: Statistical Mechanics and its Applications}} \textbf{\bibinfo{volume}{535}}, \bibinfo{pages}{122521} (\bibinfo{year}{2019}).

\bibitem{rzezonka2022attempt}
\bibinfo{author}{Rzezonka, J.}, \bibinfo{author}{Chraibi, M.}, \bibinfo{author}{Seyfried, A.}, \bibinfo{author}{Hein, B.} \& \bibinfo{author}{Schadschneider, A.}
\newblock \bibinfo{title}{An attempt to distinguish physical and socio-psychological influences on pedestrian bottleneck}.
\newblock \emph{\bibinfo{journal}{Royal Society open science}} \textbf{\bibinfo{volume}{9}}, \bibinfo{pages}{211822} (\bibinfo{year}{2022}).

\end{thebibliography}

\end{document}